\documentclass[10pt,journal,compsoc]{IEEEtran}
\usepackage{amsmath}

\usepackage{dcolumn}
\ifCLASSOPTIONcompsoc
  \usepackage[nocompress]{cite}
\else
  \usepackage{cite}
\fi
\ifCLASSINFOpdf
\else
\fi
\usepackage{graphicx}
\usepackage{amsmath}
\begin{document}
%
\title{Quantum Algorithms for Prediction Based on Ridge Regression}
%
%
%
%

\author{Menghan~Chen,~\IEEEmembership{}
    	Chaohua~Yu,~\IEEEmembership{}
       Gongde~Guo,~\IEEEmembership{}
        and~Song~Lin,~\IEEEmembership{}
\IEEEcompsocitemizethanks{\IEEEcompsocthanksitem Menghan Chen is School of Mathematics and Informatics, Fujian Normal University, Fuzhou 350007, China.\protect\\
E-mail: 1446514387@qq.com
\IEEEcompsocthanksitem Gongde~Guo and Song~Lin are with College of Mathematics and Informatics, Fujian Normal University, Fuzhou 350117, China.
 Chaohua~Yu is School of Information Management, Jiangxi University of Finance and Economics.}
\thanks{Manuscript received ; revised .}}

%
%

\markboth{Journal of \LaTeX\ Class Files,~Vol, No, }%
{Shell \MakeLowercase{\textit{et al.}}: Bare Advanced Demo of IEEEtran.cls for IEEE Computer Society Journals}
%



\IEEEtitleabstractindextext{%
\begin{abstract}
We propose a quantum algorithm based on ridge regression model, which get the optimal fitting parameters $\vec w$ and a regularization hyperparameter $\alpha$ by analysing the training dataset. The algorithm consists of two subalgorithms. One is generating predictive value for a new input, the way is to apply the phase estimation algorithm to the initial state $\left| {X} \right\rangle $ and apply the controlled rotation to the eigenvalue register. The other is finding an optimal regularization hyperparameter $\alpha$, the way is to apply the phase estimation algorithm to the initial state $\left| {\vec y} \right\rangle $ and apply the controlled rotation to the eigenvalue register. The second subalgorithm can compute the whole training dataset in parallel that improve the efficiency. Compared with the classical ridge regression algorithm, our algorithm overcome multicollinearity and overfitting. Moreover, it have exponentially faster. What's more, our algorithm can deal with the non-sparse matrices in comparison to some existing quantum algorithms and have slightly speedup than the existing quantum counterpart. At present, the quantum algorithm has a wide range of application and the proposed algorithm can be used as a subroutine of other quantum algorithms.
\end{abstract}

}

\maketitle

\IEEEdisplaynontitleabstractindextext

%
\IEEEpeerreviewmaketitle

\ifCLASSOPTIONcompsoc
\IEEEraisesectionheading{\section{Introduction}\label{sec:introduction}}
\else
\section{Introduction}
\label{sec:introduction}
\fi

%
%
%
%
\IEEEPARstart{W}{ith} the amount of stored data globally increasing by about 20$\%$ every year[1], the pressure to find efficient approaches to machine learning is also growing[2]. At present, there are some interesting ideas that some researcherss exploit the potential of quantum computing to optimize algorithms of classical computing. In the past few decades, physicists have demonstrated the astonishing capacity of quantum system to process huge datasets[3]. Compared with the physical realization of traditional computer based on “0” and “1”, quantum computer can take advantage of the superposition of quantum states to realizes the goal of acceleration. However, the laws of quantum mechanics also limit our access to information stored in quantum systems[4,5], thus, it is a great challenge to find a better quantum algorithm over classical algorithm. Even so, there are a number of well-known improved quantum algorithms have been put forward. For example, Shor demonstrated that quantum algorithm of finding the prime factors of integer and so-called ‘discrete logarithm’ problem that could solved exponentially faster than any known classical algorithm in 1994[6]. Compared with the classical algorithms, the existing quantum algorithms have amazing speedup so that some researchers are considering whether these quantum algorithms can be applied to the filed of machine learning to solve the problems of low efficiency caused by the classical computers process big data. 

As of now, a series of quantum algorithms about \textit{Linear Regression} (LR) have been put forward. For example, Wiebe et al.[7] first provided a quantum linear regression algorithm that can efficiently solve data fitting over an exponentially large dataset by building upon an algorithm for solving systems of linear equations efficiently in 2012[8], but design matrice of their algorithm must be sparse. Hence, in 2016, Schuld et al.[9] came up with a quantum linear regression algorithm that can efficiently process the low-rank non-sparse design matrices. Although many such articles have been proposed, majority of these quantum linear regression algorithms are based on \textit{Ordinary Linear Regression} (OLR) rather than \textit{Ridge Regression} (RR). However, there are two intractable problems about multicollinearity and overfitting about LR. Thus, in 2017, Liu and Zhang[10] proposed an efficient quantum algorithm that can solve and combat questions about multicollinearity and overfitting. Nonetheless, Liu and Zhang don’t determine a good $\alpha$ for RR. So later, Yu et al.[11] presented an improved quantum ridge regression algorithm that proposed the parallel Hamiltonian simulation to obtain the optimal fitting parameters $\vec w$ and quantum K-fold cross validation to obtain an appropriate $\alpha$.

On the basis of previous papers, we further exploit how to extent RR algorithm can be faster by quantum computing than classical computing. Therefore, we design a more comprehensive quantum algorithm for RR. Inspired by quantum OLR of Schuld, we want to introduce a regularization hyperparameter $\alpha$ to the paper through analyzing data suffering from multicollinearity and overfitting. It is shown that our algorithm exponentially faster than classical counterpart when matrix is low-rank, but slightly worse on the error. To a certain extent, our algorithm have improved the existing quantum ridge regression algorithms[10, 11]. For example, we have a samll improvement compared with Yu’s algorithm, the dependence on conditional number $\kappa$ is slightly better. 

The rest of this paper is organized as follows. In Section2, we review RR in terms of basic ideas and classical algorithmic procedures. In Section3, we present two important subalgorithms and their time complexity. In Section4, we analysis time complexity of two subalgorithm and whole algorithm. Conclusion is given in the Section5.

\section{Preliminaries}
In a simple linear regression model, a training set with $\emph{M}$ data points $\left( {\vec x_i ,y_i } \right)_{i = 1}^M $ is given. Here  $\overrightarrow {{x_i}}  = {\left( {{x_{i1}},{x_{i2}}, \ldots ,{x_{iN}}} \right)^T} \in {R^N}$ is a vector of $\emph{N}$ independent input variables, ${y_i} \in R$ is the scalar dependent output variables and target vector $\vec y = {\left( {{y_1},{y_2}, \ldots ,{y_M}} \right)^T} \in R^M$. The goal of LR algorithm is to obtain a vector of fitting parameters $\vec w$ from the training set, then utilize $\vec w$ to predict the result for a new input $\vec x'$. Thus, we need to make the predictive values $f\left( {{{\vec x}_i}} \right)$ as close as possible to real values ${y_i}$. Where $f\left( {{{\vec x}_i}} \right) = {\vec x_i}^T\vec w{\rm,}$. 

A direct way to obtain optimal fitting parameters $\vec w$ is utilizing least squares method of minimizing the objective function 
\begin{equation}  
\mathop {\min }\limits_w {\left\| {X\vec w - \vec y} \right\|^2.}
\label{eq:1}
\end{equation}
where $X = \left( {\vec x_1 ,\vec x_2 ,...,\vec x_M } \right)^T  \in R^{M \times N}$. Thus, we can obtain fitting parameters $\vec w$, that is 
\begin{equation}
\vec w = {\left( {X^\dag X} \right)^{ - 1}}{X^\dag}\vec y,
\label{eq:2}
\end{equation}
Here, $X^\dag$ is Hermitian conjugate or adjoint of the matrix \emph{X}. The above regression is named as OLR.

However, in Ref.[12], Hoerl and Kennard pointed that OLR can encounter multicollinearity of independent variables of data points or overfitting from Eq. (1), so it is often far from satisfaction. To overcome the two problems, they put forward a new regression model -- RR model, in which  \textit{second normal form} (2nd NF) of $\vec w$ is introduced. Thus, the objective function is changed to 
\begin{equation}                                           
\;\mathop {\min }\limits_w {\left\| {X\vec w - \vec y} \right\|^2} + \alpha {\left\| {\vec w} \right\|^2}.
\label{eq:3}
\end{equation} 
Through simple calculation, the optimal fitting parameters $\vec w$ of the RR algorithm is
\begin{equation} 
\vec w = \left( {X^\dag  X + \alpha I} \right)^{ - 1} X^\dag  \vec y,
\label{eq:4}
\end{equation}
Obviously, OLR is a particular case of RR that $\alpha  = 0$. For any matrix $\emph{X}$, we take advantage of the singular value decomposition[13] to write $X =  U\Sigma { V^\dag }$. Here $\Sigma$ is a diagonal matrix containing the real singular values ${\lambda _1},{\lambda _2}, \dots, {\lambda _R} > 0$. Without loss of generality, we assume $\lambda _r  \in \left[ {\frac{1}{\kappa },1} \right]$ ($\kappa$ is the conditional number of matrix $\emph{X}$), and the $\emph{r}$th orthogonal column of ${U} \in {R^{M \times R}}\;({V}\;\in {R^{R\times N}})$ is left (right) eigenvector $ \vec{u}_r \left( \vec{ v}_r  \right)$ corresponding to the singular value $\lambda_r$. Because the singular value decomposition can always be applied for any matrices, we let ${Z} \buildrel \Delta \over = \left( {{X}^\dag {X} + \alpha {I}} \right)^{ - 1} {X}^\dag$, which can be represented as ${Z} = {V}\Lambda {U}^\dag  $ by the singular value decomposition, where $\Lambda $ is a diagonal matrix containing the singular values ${\sigma _1} = \frac{{{\lambda _1}}}{{\lambda _1^2 + \alpha }},{\sigma _2} = \frac{{{\lambda _2}}}{{\lambda _2^2 + \alpha }},\dots,{\sigma _R} = \frac{{{\lambda _R}}}{{\lambda _R^2 + \alpha }}$.

The following is the using of this mathematical background, $\emph{X}$ and $\emph{Z}$ have alternative expression formula 
$X = \sum\limits_r {\lambda _r } \vec u_r \vec v_r^T  $ and $Z = \sum\limits_r {\sigma _r } \vec v_r \vec u_r^T  $. So Eq. (4) can express as $\vec{w} = \sum\limits_{r = 1}^R {\sigma _r \vec{v}_r \vec{u}_r^T \vec{y}}$.
According to $f\left( {\vec{x}_i } \right) = \vec{x}_i \vec {w}$, the corresponding output $y'$ can be gained as follows for a new input $\vec x'$, 
\begin{equation}
y' = \sum\nolimits_{r = 1}^R {\sigma _r \vec x'} \vec v_r \vec u_r ^T \vec y.
\label{eq:5}
\end{equation}
The desired output is a single scalar value, but we need to get a vector for whole training set in algorithm 2. Thus, Eq. (5) must be performed many times. If so, the efficiency of algorithm is low in classical algorithm. Therefore, this work describe a quantum algorithm that can efficiently reproduce the whole result of a training dataset through parallelism of quantum computing.   

\section{Quantum algorithms for ridge regression}
\label{sec3}
In this section, we design a quantum algorithm for RR. It contains two parts: a quantum algorithm for generating predictive value $y'$ about a new input $\vec x'$ and a quantum algorithm for finding an optimal regularization hyperparameter $\alpha$. 

\subsection{Algorithm 1: generating predictive values }
\subsubsection{The specific steps of algorithm 1}

Given a training set of $\emph{M}$ data points $\left( {\vec {x}_i ,y_i } \right)_{i = 1}^M $, then we present a quantum algorithm to generate predictive values $y'$ for a new input $\vec x'$ that approximates to the real value $y$ within error $\varepsilon$. The algorithm 1 proceeds as following steps and the schematic is given in Fig. 1. 

\begin{figure*}[htbp]
\includegraphics[width=1\textwidth]{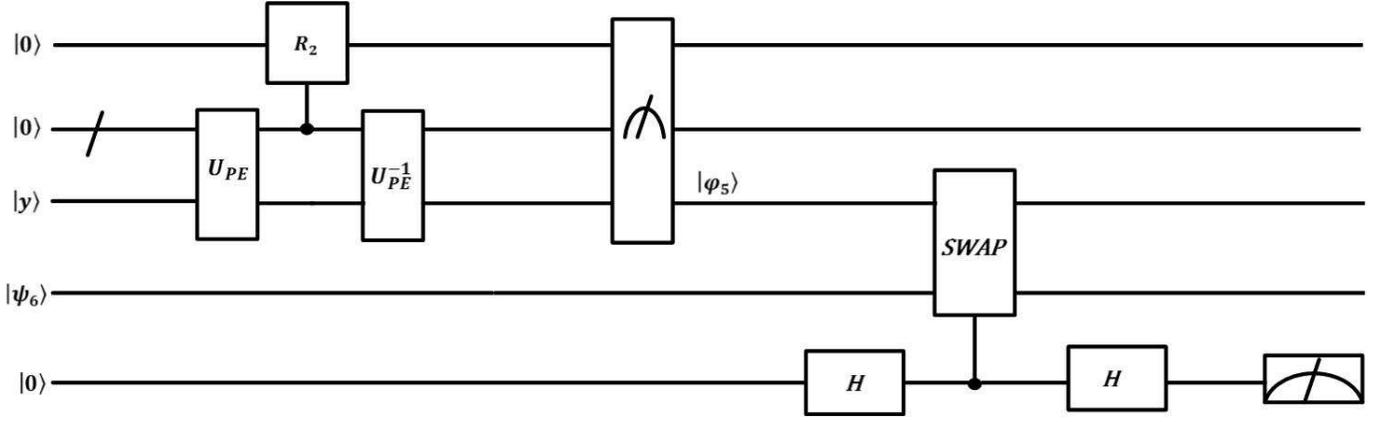}
\caption{\label{fig:epsart} The whole quantum circuit for algorithm 1. Here the $U_{PE}$ and $U_{PE}^{ - 1}$ denote the phase estimation, the inverse phase estimation of the Step (2) and Step (3) respectively. The $R_1$ is controlled rotation of Step (3). SWAP is the original swap operation of Step (5).}
\end{figure*}

Step (1) Preparing quantum state of $\vec{y}$, $\vec x'$  and $\emph{X}$ respectively. The specific expressions are 
\begin{equation}
\left|\vec{y} \right\rangle _1  = \sum\limits_{i = 0}^{M - 1} {y_i\left| i \right\rangle _1 }.
\label{eq:6}
\end{equation}

\begin{equation}
\left| \vec x' \right\rangle _2  = \sum\limits_{\gamma  = 0}^{N - 1} {{x'}_\gamma  \left| \gamma  \right\rangle _2 }, 
\label{eq:7}
\end{equation}

\begin{equation}
\left| {X} \right\rangle _{34} = \sum\nolimits_{m = 0}^{M - 1} {\sum\nolimits_{n = 0}^{N-1} {x_{mn} \left| m \right\rangle _3 \left| n \right\rangle _4} }, 
\label{eq:8}
\end{equation}
According to reference[14], the procedure of preparing quantum state $\left|\vec{y} \right\rangle _1$ is as follows. Supposed, there is an oracle $O_{\vec {y}}$ that can access the elements of $\vec {y}$ in time $O(\log_2 {\rm{M}})$ from quantum random access memory (QRAM) and acts as 
\begin{equation}
O_{{y}} :\left| i \right\rangle_{1} \left| 0 \right\rangle_{a2}  \to \left| i \right\rangle_{1} \left| {y_i } \right\rangle_{a2}. 
\label{eq:9}
\end{equation}
Secondly, giving initial state $\left| 0 \right\rangle_{1} \left| 0 \right\rangle_{a2}$ and performing the quantum Fourier transform on $\left| 0 \right\rangle_{1}$, which gets $\sum\limits_{i = 0}^{M - 1} {\frac{{\left| i \right\rangle_{1} }}{{\sqrt M }}} $. The quantum Fourier transform  on an orthonormal basis $\left| 0 \right\rangle ,...,\left| {N - 1} \right\rangle $ is defined to be a linear operator with the following action on the basis states, 
\begin{equation}
\left| j \right\rangle  \to \frac{1}{{\sqrt N }}\sum\limits_{k = 0}^{N - 1} {{e^{{{2\pi ijk} \mathord{\left/
					{\vphantom {{2\pi ijk} N}} \right.
					\kern-\nulldelimiterspace} N}}}\left| k \right\rangle } .
\label{eq:10}
\end{equation}
Then, the oracle is applied to  $\sum\limits_{i = 0}^{M - 1} {\frac{{\left| i \right\rangle_{1} \left| 0 \right\rangle_{a2} }}{{\sqrt M }}} $, which come into being $\sum\limits_{i = 0}^{M - 1} {\frac{{\left| i \right\rangle_{1} \left| {y_i } \right\rangle_{a2} }}{{\sqrt M }}}$. Thirdly, appending a qubit $\left| 0 \right\rangle_{a3}$ and performing controlled rotation for $\left| {y_i } \right\rangle_{a2}$ to generate the state 
\begin{equation}
\sum\limits_{i = 0}^{M - 1} {\frac{{\left| i \right\rangle_{1} \left| {y_i } \right\rangle_{a2} }}{{\sqrt M }}} \left( {\sqrt {1 - \left( {\frac{{y_i }}{{\left\| {\vec {y}} \right\|_{\max } }}} \right)^2 } \left| 0 \right\rangle_{a3}  + \frac{{y_i }}{{\left\| {\vec {y}} \right\|_{\max } }}\left| 1 \right\rangle_{a3} } \right).
\label{eq:11}
\end{equation}
In the next moment, uncomputing the oracle and generating the state
\begin{equation}
\sum\limits_{i = 0}^{M - 1} {\frac{{\left| i \right\rangle_{1} \left| 0 \right\rangle_{a2} }}{{\sqrt M }}\left( {\sqrt {1 - {{\left( {\frac{{{y_i}}}{{\left\| {\vec y} \right\|}}} \right)}^2}} \left| 0 \right\rangle_{a3}  + \frac{{{y_i}}}{{\left\| {\vec y} \right\|}}\left| 1 \right\rangle_{a3} } \right)}.
\label{eq:12}
\end{equation}
Finally, measuring the last register a3 in the basis $\left\{ {\left| 0 \right\rangle ,\left| 1 \right\rangle } \right\}$, which has probability $P_{y}  = \frac{{\sum\limits_{i = 0}^{M - 1} {y_i^2 } }}{{M\left\| {\vec{y}} \right\|_{\max }^2 }}$.  Generally, we assume that $\vec {y}$ is balanced ($\vec x'$ and $\emph{X}$ is also same), so  $P_{y}  = \Omega \left( 1 \right)$. This implies that we need $O\left( 1 \right)$ measurements to obtain $\sum\limits_{i = 0}^{M - 1} {\frac{{y_i\left| i \right\rangle_1 }}{{\sqrt M \left\| \vec {y} \right\|_{\max } }}}$. That is, we have a large probability to get the desired result. Moreover, the total time for generating the the results is $O(\log_2 {\rm{M}})$.
Thus, we can easily construct the quantum state $\left|\vec{y} \right\rangle _1$
\begin{equation}
\left|\vec {y} \right\rangle _1  = \sum\limits_{i = 0}^{M - 1} {y_i\left| i \right\rangle _1 }. 
\label{eq:13}
\end{equation}
            
The same method is used to get the other two quantum states respectively.
\begin{equation}
\left| \vec x' \right\rangle _2  = \sum\limits_{\gamma  = 0}^{N - 1} {{x'}_\gamma  \left| \gamma  \right\rangle _2 }, 
\label{eq:14}
\end{equation}
\begin{equation} 
\left| {X} \right\rangle _{34} = \sum\nolimits_{m = 0}^{M - 1} {\sum\nolimits_{n = 0}^{N-1} {x_{mn} \left| m \right\rangle _3 \left| n \right\rangle _4} }, 
\label{eq:15}
\end{equation}
where $\sum\limits_i {{{\left| {{y_i}} \right|}^2}} = \sum\limits_\gamma  {\left| {{x'}_\gamma  } \right|} ^2 = 
\sum\limits_{m,n} {{{\left| {{x_{mn}}} \right|}^2}}  =  1$. Then using the Gram-Schmidt decomposition[15], $\left|{X} \right\rangle _{34}$ can be re-expressed as
\begin{equation}
\left| {X} \right\rangle _{34} {\rm{ = }}\sum\limits_{r = 1}^R {\lambda _r } \left| {\vec {u}_r } \right\rangle _3 \left| {\vec {v}_r } \right\rangle _4 .
\label{eq:16}
\end{equation}
Here $\left| {{\vec {u}_r}} \right\rangle _3 $ and $\left| {{\vec {v}_r}} \right\rangle _4 $ are quantum states representing the orthogonal sets of left and right eigenvectors of $X$ respectively. 

Step (2) In order to convert Eq. (16) to “quantum representation” of result (5), that is to say, quantum form of $y' = \sum\nolimits_{r = 1}^R {\sigma _r \vec x'} \vec {v}_r \vec {u}_r ^T \vec {y}$. We are first to extract the singular value $\lambda_r^2$ of $\emph{X}{\emph{X}^\dag }$. Since $\emph{X}$ is generally not Hermitian, we extent it to a Hermitian matrix $\emph{X}{\emph{X}^\dag }$. Then, we need to calculate the reduced density operator for the register $\left| {\vec {u}_r } \right\rangle _3$ of $\left| {X} \right\rangle _{34}$. So in that case, we can get a mixed state $\rho _{ {X} {X}^\dag }$. 
\begin{equation}
\rho _{{X}{{X}^\dag }}  = tr_4 \left\{ {\left| {X} \right\rangle \left\langle {X} \right|} \right\} = \sum\nolimits_{r = 1}^R {\lambda _r^2 \left| {\vec {u}_r } \right\rangle } \left\langle {\vec {u}_r } \right|,
\label{eq:17}
\end{equation}
where $tr_4$ is known as the partical trace over the register $\left| {\vec {v}_r } \right\rangle _4$. Here, we take the trick of reference[16] to apply $e^{ - i\rho _{ {X}{{X}^\dag } } q\Delta t}$ to 
$\left| {X} \right\rangle _{34}$ resulting in 
\begin{equation}
\left| \varphi  \right\rangle _1  = \sum\nolimits_{q = 1}^Q {\left| {q\Delta t} \right\rangle } \left\langle {q\Delta t} \right|e^{ - i\rho _{{X}{{X}^\dag }} q\Delta t} \left| {X} \right\rangle \left\langle {X} \right|e^{i\rho _{{X}{{X}^\dag } } q\Delta t},
\label{eq:18}
\end{equation}
for some large \emph{Q}. Then, by performing the quantum phase estimation algorithm on 
$\left| {X} \right\rangle _{34}$, we can get
\begin{equation}
\left| {\varphi _2 } \right\rangle _{345}  = \sum\nolimits_{r = 1}^R {\lambda _r } \left| {\vec {u}_r } \right\rangle _3 \left| {\vec {v}_r } \right\rangle _4 \left| {\lambda _r^2 } \right\rangle _5.
\label{eq:19}
\end{equation}

The phase estimation is performed in two stages. First, we run the quantum circuit shown in Fig. 2. The circuit begins by applying the quantum Fourier transform to the first register $\left| 0 \right\rangle  $, followed by application of controlled unitary operation on the second register $\left| {X} \right\rangle _{34}$ that unitary operator is $e^{ - i\rho _{{X}{{X}^\dag } } t} $.

\begin{figure}
   \centering
    \includegraphics [width=0.5\textwidth] {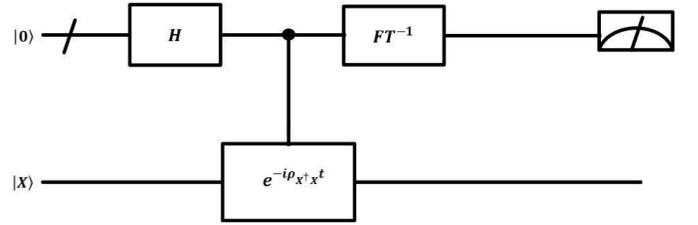}
     \centering
    \caption{Schematic of the overall phase estimation procedure. The top multiple qubits are the first register(the '/'  usually denote a bundle of wires), and the bottom qubits are second register. The $FT^{ - 1}$ refers to the quantum Fourier transformation.}
   \label{Fig:2}
  \end{figure}

Step (3) Adding an ancilla qubit $\left| 0 \right\rangle _6$, then rotating it from $\left| 0 \right\rangle _6$ to 
$\sqrt {1 - \left( {\frac{{C_1 }}{{\lambda _r^2  + \alpha }}} \right)^2 } \left| 0 \right\rangle _6 + \frac{{C_1 }}{{\lambda _r^2  + \alpha }}\left| 1 \right\rangle _6
$ conditioned by 
$\left| {\lambda _r^2} \right\rangle _5$. In this way, we can obtain
\begin{equation}
 \begin{split}
\left| {\varphi _3 } \right\rangle _{3456} = 
& \sum\nolimits_{r = 1}^R {\lambda _r } \left| {\vec {u}_r } \right\rangle _3 \left| {\vec {v}_r } \right\rangle _4 \left| {\lambda _r^2 } \right\rangle _5 \\
&\left( {\sqrt {1 - \left( {\frac{{C_1 }}{{\lambda _r^2  + \alpha }}} \right)^2 } \left| 0 \right\rangle _6  + \frac{{C_1 }}{{\lambda _r^2  + \alpha }}\left| 1 \right\rangle _6 } \right), 
\label{eq:20}
\end{split}
\end{equation}
where the constant ${C_1}$ is chosen to be $O\left( {\max \left( {\frac{{\lambda _r }}{{\lambda _r^2  + \alpha }}} \right)} \right)^{ - 1} $ that makes $\frac{{C_1 }}{{\lambda _r^2  + \alpha }}$ as close to 1 as possible while less than 1. The $\alpha $ satisfying $\left[ {\Theta \left( {\frac{1}{{\kappa ^2 }}} \right),\Theta \left( 1 \right)} \right]$ (it's explained later). Then, after performing quantum inverse phase estimation algorithm and discarding the third register $\left| {\lambda _r^2 } \right\rangle _3$, the remainder particles are in the state 
\begin{equation}
 \begin{split}
\left| {\varphi _4 } \right\rangle _{346} = 
& \sum\nolimits_{r = 1}^R {\lambda _r } \left| {\vec {u}_r } \right\rangle _3 \left| {\vec {v}_r } \right\rangle _4  \\
&\left( {\sqrt {1 - \left( {\frac{{C_1 }}{{\lambda _r^2  + \alpha }}} \right)^2 } \left| 0 \right\rangle _6 + \frac{{C_1 }}{{\lambda _r^2  + \alpha }}\left| 1 \right\rangle _6 } \right).
\label{eq:}
\end{split}
\end{equation}

Step (4) A projection measurement is performed on the ancilla qubit in the basis 
$\left\{ {\left| 0 \right\rangle ,\left| 1 \right\rangle } \right\}$. Then the probability of obtaining measurement outcome 
${\left| 1 \right\rangle _6}$ is $p\left( 1 \right) = \sum\limits_{r = 1}^R {\lambda _r^2\frac{{{C_1}^2}}{{{{\left( {\lambda _r^2 + \alpha } \right)}^2}}}} $, from  $\frac{{C_1 }}{{\lambda _r^2  + \alpha }} = \Omega \left( {\frac{1}{\kappa }} \right)$, we can derive
$p\left( 1 \right) = O\left( {\frac{1}{{{k^4}}}} \right)$.
After the measurement, the remainder system is in the state
\begin{equation}
\left| {\varphi _5 } \right\rangle _{34}  =  \frac{{\sum\limits_{r = 1}^R {{\lambda _r}} \left| {{\vec {u}_r}} \right\rangle _3\left| {{\vec {v}_r}} \right\rangle _4 \frac{{{C_1}}}{{\lambda _r^2 + \alpha }}}}{{\sqrt {p\left( 1 \right)} }}.
\label{eq:22}
\end{equation}

Step (5) 
The aim of the last step is to obtain the desired result. According to the $\left| {\varphi _5 } \right\rangle _{34}$, $\left| {{\vec {y} }} \right\rangle _1$ and $\left| {\vec x'} \right\rangle _2$, we can get predictive output ${y'}$ by the following procedures. First, we need to express the desired result in quantum states in order to make the calculation clearer
\begin{equation} 
y' = \sum\limits_{r = 1}^R {\sigma _r \left\langle {{\vec {u}_r }}
	\mathrel{\left | {\vphantom {{u_r } y}}
		\right. \kern-\nulldelimiterspace}
	\vec {y} \right\rangle \left\langle {{\vec {v}_r }}
	\mathrel{\left | {\vphantom {{\vec {v}_r } {\vec x'}}}
		\right. \kern-\nulldelimiterspace}
	{{\vec x'}} \right\rangle }. 
\label{eq:23}
\end{equation}
Secondly, observing state (23) and defining $\left| {\varphi } \right\rangle _{12}  = \left| \vec {y} \right\rangle _1 \left| {\vec x'} \right\rangle _2 $ according to Eq. (6) and Eq. (7). The result of performing inner product on $\left| {\varphi _5 } \right\rangle _{34}$ and $\left| {\varphi _6 } \right\rangle _{12}$ is in direct proportion to our desired outcome by a Swap Test[17, 18], which is performing Hadamard transform on ancilla qubit ${\left| 0 \right\rangle_0 }$, then executing swap operation on $\left| {\varphi _5 } \right\rangle _{34}$ and $\left| {\varphi _6 } \right\rangle _{12}$ when the ancilla qubit is 
${\left| 1 \right\rangle }$, followed by measurement on the ancilla qubit in the basis $\left\{ {\left| 0 \right\rangle ,\left| 1 \right\rangle} \right\}$. But the direct operation will result in the probability of success is 
$\frac{1}{2} + \frac{1}{2}\left| {\left\langle {{\varphi _5 }}
 \mathrel{\left | {\vphantom {{\varphi _5 } {\varphi _6 }}}
 \right. \kern-\nulldelimiterspace}
 {{\varphi _6 }} \right\rangle } \right|^2 $, it is easy to see that the sign of ${\left\langle {{\varphi _5 }}\mathrel{\left | {\vphantom {{\varphi _5 } {\varphi _6 }}}\right.\kern-\nulldelimiterspace}
 {{\varphi _6 }} \right\rangle }$ is ambiguous. Therefore, there is a more deliberate way can avoid the terrible situation: Conditionally preparing these two states to make them entangle with ancilla qubit[19], that is
$\frac{{\left| 0 \right\rangle_0 \left| {\varphi _5 } \right\rangle _{34} + \left| 1 \right\rangle_0 \left| {\varphi _6 } \right\rangle _{12} }}{{\sqrt 2 }}
$ and then performing the Swap Test on the ancilla qubit with 
$\frac{{\left| 0 \right\rangle  - \left| 1 \right\rangle }}{{\sqrt 2 }}$. The probability of success is 
$Pr = \frac{1}{4} + \frac{1}{4}\left\langle {{\varphi _5 }}
 \mathrel{\left | {\vphantom {{\varphi _5 } {\varphi _6 }}}
 \right. \kern-\nulldelimiterspace}
 {{\varphi _6 }} \right\rangle $ that can lead to reveal sign. The circuit of changed Swap Test is given in Fig. 3. In practice, a good RR model make predictive value as close as possible real value ${y_i}$, thus the sum of inner product always positive in this case.

\begin{figure}
    \centering
    \includegraphics [width=0.5\textwidth] {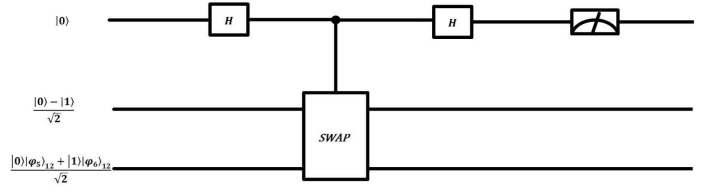}
    \caption{The quantum circuit for the changed Swap Test. The  \emph{H} is Hadamard operation and \emph{SWAP} is swap operation.}
    \label{Fig:3}
  \end{figure}

\subsection{Algorithm 2: finding an optimal $\alpha $} 
\subsubsection{The specific steps of algorithm 2}
According to reference[11,18], we know that a good $\alpha$ can make RR model achieve the best (or approximately best) predictive performance. So, it is of great significance to choose a good $\alpha$.

For RR model, the reference[11] proposed that too large $\alpha $ will make the optimal fitting parameters $\vec w$ approach zero which leads to deviate real value and too small $\alpha $ will make the RR reduced to the OLR. Thus, we choose $\alpha $ satisfying $\left[ {\Theta \left( {\frac{1}{{\kappa ^2 }}} \right),\Theta \left( 1 \right)} \right]$ based on range of the singular value $\lambda _r$. That is to say, ${\alpha _j} = {\alpha _{min}} + \frac{{\left( {j - 1} \right)\left( {{\alpha _{max}} - {\alpha _{min}}} \right)}}{{L - 1}}$ for $j = 1,\dots,L$. The common way to get an appropriate $\alpha $ is to choose the best one out of a number of candidates $\alpha '$ . The goal of the following steps is to find an appropriate $\alpha $ that give rise to a RR model can well constructed. 

For every given input $\vec {x}_i $, we will get a predictive value ${y_i}' $. However, there exist the squared residual sum between predictive values and real values. The predictive values are related to $\alpha $, so we hope that we can find an optimal $\alpha $ to obtain minimum of the squared residual sum. In the algorithm 1, our algorithm only can obtain an output value every time. So, the efficiency is low if we want to input whole training dataset $\emph{X}$. Therefore, we propose another subalgorithm that can calculate training dataset $\emph{X}$ in parallel. 

Here, we write ${X}$ in the reduced singular value decomposition form and combine it with Eq. (5), then, we can get a set of column vectors corresponds to $\emph{M}$ outputs. Where the desired quantum state form $\vec y'$ is 
\begin{equation}
\left| {\vec x'} \right\rangle  _1 = \sum\limits_{r = 1}^R {\frac{{\lambda _r^2}}{{\lambda _r^2 + \alpha }}} \left\langle {{{\vec {u}_r}}}
\mathrel{\left | {\vphantom {{{\vec {u}_r}} y}}
	\right. \kern-\nulldelimiterspace}
{\vec {y}} \right\rangle \left| {{\vec {u}_r}} \right\rangle _1.
\label{eq:18}
\end{equation}

The details of the second algorithm are described in the following steps and the schematic circuit is given in Fig. 4.
\begin{figure*}[htbp]
	\includegraphics[width=1\textwidth]{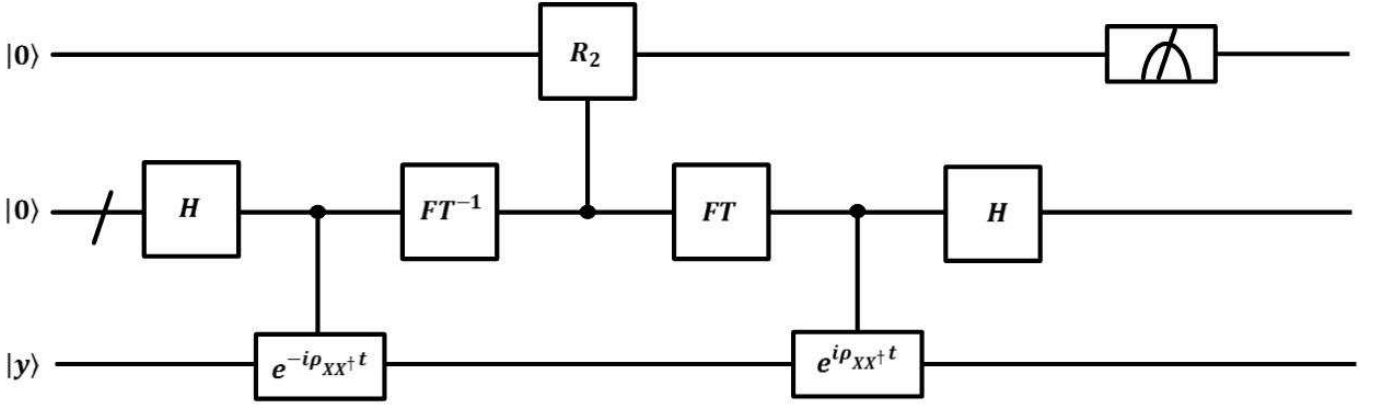}
	\caption{\label{fig:epsart} The whole quantum circuit for algorithm 2. Here the '/'  usually denote a bundle of wires, $H$ refers to Hadamard operation. The $FT$ means quantum Fourier transformation and $FT^{ - 1}$ denotes the quantum inverse Fourier transformation. The $R_2$ is controlled rotation of Step (3).}
\end{figure*}

Step (1) Similar to the algorithm 1, fisrtly preparing initial quantum state of $\vec{y}$, $\vec x'$  and $\emph{X}$ which can be efficiently generated as shown in Step (1) of algorithm 1. Next, taking advantage of the Gram-Schmidt decomposition to write $\left|{X} \right\rangle _{34} = \sum\limits_{r = 1}^R {{\lambda _r}} \left| {{\vec {u}_r}} \right\rangle _3 \left| {{\vec {v}_r}} \right\rangle _4$.

Step (2) Observing Eq. (24), we need to estimate the singular value ${\lambda _r^2 }$ of ${ {X}{{X}^\dag }}$. Next, adopting trick of reference[16] to use $\emph{q}$ copies of ${\rho _{{X}{{X}^\dag } } }$ to apply the unitary operator $e^{i\rho _{{X}{{X}^\dag }} t}$, which allows us to exponentiate a non-sparse but low-rank $M \times M$ - dimension density matrices, where ${\rho _{ {X}{{X}^\dag } } }$ is presented in algorithm 1. Followed by performing the quantum phase estimation algorithm on $\left| \vec {y} \right\rangle _1$ that lead to obtain
\begin{equation}
\left| {\phi _1 } \right\rangle _{12}  = \sum\nolimits_{r = 1}^R {\left\langle {{\vec {u}_r }}
	\mathrel{\left | {\vphantom {{\vec {u}_r } y}}
		\right. \kern-\nulldelimiterspace}
	{\vec {y}} \right\rangle } \left| {\vec {u}_r } \right\rangle _1 \left| {\lambda _r^2 } \right\rangle _2.
\label{eq:18}
\end{equation}
The procedures of the phase estimation are same as counterpart of algorithm 1. Just the objective of application by ${\rho _{ {X}{{X}^\dag } }}$ is from $\left| {X} \right\rangle _{34}$ to $\left| \vec {y} \right\rangle _1$.	 	

Step (3) Adding an ancilla qubit $\left| 0 \right\rangle _3$, then rotating it from $\left| 0 \right\rangle _3$
to $
\sqrt {1 - \left( {\frac{{C_2 \lambda _r^2 }}{{\lambda _r^2  + \alpha }}} \right)^2 } \left| 0 \right\rangle _3  + \frac{{C_2 \lambda _r^2 }}{{\lambda _r^2  + \alpha }}\left| 1 \right\rangle _3$ conditioned on 
${\left| {\lambda _r^2} \right\rangle _2}$. In this way, we can obtain
\begin{equation}
\begin{split}
\left| {\phi _2 } \right\rangle _{123}  =
& \sum\nolimits_{r = 1}^R {\left\langle {{\vec {u}_r }}
	\mathrel{\left | {\vphantom {{\vec {u}_r } \vec {y}}}
		\right. \kern-\nulldelimiterspace}
	\vec {y} \right\rangle } \left| {\vec {u}_r } \right\rangle _1 \left| {\lambda _r^2 } \right\rangle _2 \\
&\left[ {\sqrt {1 - \left( {\frac{{C_2 \lambda _r^2 }}{{\lambda _r^2  + \alpha }}} \right)^2 } \left| 0 \right\rangle _3  + \frac{{C_2 \lambda _r^2 }}{{\lambda _r^2  + \alpha }}\left| 1 \right\rangle _3 } \right],
\label{eq:}
\end{split}
\end{equation}
where the constant  ${C_2}$ is chosen to be $O\left( 1 \right)$ for ensuring that amplitude as close as possible to 1 but less than 1. In the next moment, performing the the quantum inverse phase estimation algorithm and discarding the eigenvalue register $\left| {\lambda _r^2 } \right\rangle _2$. After these operations, the remainder system is in the state
\begin{equation}
\begin{split}
\left| {\phi _3 } \right\rangle _{13}  =
& \sum\nolimits_{r = 1}^R {\left\langle {{\vec {u}_r }}
	\mathrel{\left | {\vphantom {{\vec {u}_r } \vec {y}}}
		\right. \kern-\nulldelimiterspace}
	\vec {y} \right\rangle } \left| {\vec {u}_r } \right\rangle _1 \\
&\left[ {\sqrt {1 - \left( {\frac{{C_2 \lambda _r^2 }}{{\lambda _r^2  + \alpha }}} \right)^2 } \left| 0 \right\rangle _3  + \frac{{C_2 \lambda _r^2 }}{{\lambda _r^2  + \alpha }}\left| 1 \right\rangle _3 } \right].
\label{eq:20}
\end{split}
\end{equation}

Step (4) Performing conditional measurement on the ancilla qubit in the basis $\left\{ {\left| 0 \right\rangle ,\left| 1 \right\rangle } \right\}$ to get outcome 
$\left| 1 \right\rangle _3$ and final state after measurement is
\begin{equation}
\left| {\phi _4 } \right\rangle _1 
= \frac{{\sum\limits_{r = 1}^R {\frac{{{C_2}\lambda _r^2}}{{\lambda _r^2 + \alpha }}\left\langle {{{\vec {u}_r}}}
			\mathrel{\left | {\vphantom {{{\vec {u}_r}} \vec {y}}}
				\right. \kern-\nulldelimiterspace}
			\vec {y} \right\rangle \left| {{\vec {u}_r}} \right\rangle _1 } }}{{\sqrt {p\left( 2 \right)} }},
\label{eq:21}
\end{equation}
with 
$p\left( 2 \right) = {\left( {\sum\limits_{r = 1}^R {\left\langle {{{\vec {u}_r}}}
			\mathrel{\left | {\vphantom {{{\vec {u}_r}} \vec {y}}}
				\right. \kern-\nulldelimiterspace}
			\vec {y} \right\rangle } \frac{{{C_2}\lambda _r^2}}{{\lambda _r^2 + \alpha }}} \right)^2}$. It is easy to see that  
$\left| {\vec y'} \right\rangle _1 = \frac{{\sqrt {p(2)} \left| {\phi _4 } \right\rangle _1 }}{{{C_2}}}$. 

Step (5) Given a set of candidates $\left\{ {\alpha _1 ,...,\alpha _L } \right\}$, computing the loss function
\begin{equation}
\begin{split}
E\left( {\alpha _l } \right) &= \left| {\left| {\vec y'} \right\rangle  - \left| \vec {y} \right\rangle _1} \right|^2 \\ 
& =\left\langle {\vec y'|{\vec y'}} \right\rangle  + \left\langle {\vec {y}|\vec {y}} \right\rangle  - 2\left\langle {\vec y'|\vec {y}} \right\rangle\\  
&=\frac{{p\left( 2 \right)}}{{C_2^2}} + 1 - 2\frac{{\sqrt {p(2)} }}{{{C_2}}}\left\langle {{\phi _3 }}
\mathrel{\left | {\vphantom {{\phi _4 } y}}
	\right. \kern-\nulldelimiterspace}
\vec {y} \right\rangle, 
\label{eq:22}
\end{split}
\end{equation}
where
$p\left( 2 \right)$ is probability of measured result and 
${C_2}$ is a known constant. The step is important, because we want the fitted model to capture the typical relationship among the real values and predictive values so that it can be generalized to new data. This require that we choose  an optimal ${\hat \alpha }$ that obtain minimum $E\left( {\hat \alpha } \right)$. For $\left\langle {{\phi _4 }}
\mathrel{\left | {\vphantom {{\phi _4 } \vec {y}}}
	\right. \kern-\nulldelimiterspace}
\vec {y} \right\rangle$, we can adopt the Swap Test to obtain the desired consequences. 

Step (6) For every $\alpha  \in \left\{ {\alpha _1 ,\alpha _2 ,...,\alpha _L } \right\}$, executing Steps (1)-(5), then picking out the best $\alpha$ with minimum $E\left( \alpha  \right)$ as the final regularization hyperparameter ${\hat \alpha }$ for RR. Thus, we will get $\emph{L}$ results of the loss function by $\emph{L}$ repetitions of algorithm 2.

\section{Runtime analysis of algorithm}
\subsection{Runtime analysis of algorithm 1}
Density matrix exponentiation is a powerful tool to analyze the properties of unknown density matrices. According to[16], one needs time resource $O\left( {{\varepsilon ^{ - 3}}} \right)$ copies of ${\rho _{{X}{{X}^\dag }}}$ to apply the unitary operator ${e^{ - i\;{\rho _{{X}{{X}^\dag }}} t}}$, which allows accuracy 
 $\varepsilon  = O\left( {{{{t^2}} \mathord{\left/{\vphantom {{{t^2}} q}} \right.\kern-\nulldelimiterspace} q}} \right)$. Therefore, one constructs the eigenvectors and eigenvalues of a low-rank $\left( {M \times M} \right)$-dimension matrix ${\rho _{{X}{{X}^\dag }}}$ in time $O\left( {\log_2 M} \right)$. And the runtime complexity of the controlled rotation is 
$O\left( {log\left( {\frac{1}{\varepsilon }} \right)} \right)$. Compared with the phase estimation algorithm, the runtime of this step is negligible. The probability of getting predictive value is $p\left( 1 \right) = O\left( {\frac{1}{{{k^4}}}} \right)$, which means we need  $O\left( {{k^4}} \right)$ to obtain desired results on average. Amplitude amplification[20,21] reduces this times to $O\left( {{k^2}} \right)$ in the runtime. The Swap Test routine is also linear in the number of qubits, and the final measurement only accounts for a constant factor. Of course, time complexity on state preparing is $O\left( {\log_2 MN} \right)$. Thus, the upper bound of runtime roughly is $O\left( {\kappa ^2 poly\log_2 \left( {MN} \right)\varepsilon ^{ - 3} } \right)$. Comparisons of time complexity between classical counterpart, Liu and Zhang’s quantum algorithm, Yu’s improved quantum algorithm and our proposed quantum algorithm are detailed in TABLE1.

\begin{table}
\caption{Comparisons of time complexity about algorithm 1}
\begin{tabular}{cc}
\hline
\hline
Algorithm     &     Time Complexity\\
\hline
Classical counterpart & \ $ O\left( {MN + {{N^2 R\log \left( {\frac{R}{\varepsilon }} \right)} \mathord{\left/
 {\vphantom {{N^2 R\log \left( {\frac{R}{\varepsilon }} \right)} {\varepsilon ^2 }}} \right.
 \kern-\nulldelimiterspace} {\varepsilon ^2 }}} \right)
$\\
Liu's algorithm & $O\left( {\log_2 \left( {M + N} \right){{s^2 \kappa _R^3 } \mathord{\left/
 {\vphantom {{s^2 \kappa _R^3 } {\varepsilon ^2 }}} \right.
 \kern-\nulldelimiterspace} {\varepsilon ^2 }}} \right)
$\\
Yu’s algorithm &  $O\left( {\left\| {X} \right\|_{\max }^2 poly\log_2 \left( {M + N} \right){{\kappa ^3 } \mathord{\left/
 {\vphantom {{\kappa ^3 } {\varepsilon ^3 }}} \right.
 \kern-\nulldelimiterspace} {\varepsilon ^3 }}} \right)
$ \\
Our algorithm & $O\left( {{{\kappa ^2 \log M} \mathord{\left/
			{\vphantom {{\kappa ^2 \log_2 M} {\varepsilon ^3 }}} \right.
			\kern-\nulldelimiterspace} {\varepsilon ^3 }}} \right)$\\
\hline
\hline
\end{tabular}
\end{table}

From above TABLE 1, we have an exponential speedup on $\emph{MN}$ when compared with result of the best classical counterpart,  whereas the dependence on accuracy of a factor $\varepsilon $ that is small. So our algorithm 1 is accelerated totally. What’s more, there are slightly advantage when compared our algorithm with similar quantum algorithm. Because Liu and Yu didn't take state preparing into account, our time complexity is $O\left( {{{\kappa ^2 \log_2 M} \mathord{\left/
			{\vphantom {{\kappa ^2 \log M} {\varepsilon ^3 }}} \right.
			\kern-\nulldelimiterspace} {\varepsilon ^3 }}} \right)$ when we discard time complexity of state preparing.
Frist, Liu’s quantum RR algorithm, whose time complexity is 
$O\left( {\log_2 \left( {M + N} \right){{s^2 \kappa _R^3 } \mathord{\left/
 {\vphantom {{s^2 \kappa _R^3 } {\varepsilon ^2 }}} \right.
 \kern-\nulldelimiterspace} {\varepsilon ^2 }}} \right)$, where $\emph{s}$ is the sparsity of the matrix, the dependence on conditional number $\kappa $ is slightly better than his algorithm, whereas dependence on $\varepsilon $ is slightly worse. Secondly, our algorithm can deal with the non-sparse matrix. And runtime of Yu’s quantum RR is $O\left( {\left\| {X} \right\|_{\max }^2 poly\log_2 \left( {M + N} \right){{\kappa ^3 } \mathord{\left/
 {\vphantom {{\kappa ^3 } {\varepsilon ^3 }}} \right.
 \kern-\nulldelimiterspace} {\varepsilon ^3 }}} \right)$ when the matrix is low-rank. So we have a little advantages about depending on conditional number $\kappa $, and have same dependence on $\varepsilon$ as Yu’ result.   

\subsection{ Runtime analysis of algorithm 2}
Time complexity of algorithm 2 is similar to algorithm 1, just the probability of measurement is $p\left( 2 \right)$, which means we need 
$\frac{1}{{p\left( 2 \right)}}$ to obtain desired results on average. Amplitude amplification reduces this times to 
$\frac{1}{{\sqrt {p(2)} }}$ in the runtime. Here, ${\left| {{\vec {u}_r}} \right\rangle }$ are quantum states representing the orthogonal sets of left eigenvectors of ${X}$. Adding another ${M - R}$ normalized vectors 
$\left| {{\vec {u}_{R + 1}}} \right\rangle , \ldots ,\left| {{\vec {u}_M}} \right\rangle $
that make $\left| {{\vec {u}_1}} \right\rangle , \ldots ,\left| {{\vec {u}_M}} \right\rangle $ become an orthonormal basis in whole space 
${R^M}$. Therefore, $\left| \vec {y} \right\rangle $ can be written as a linear combination of
$\left\{ {\left. {\left| {{\vec {u}_{\rm{r}}}} \right\rangle } \right\}} \right._1^M
$, that is 
$\left| \vec {y} \right\rangle  _2= \sum\limits_{r = 1}^N {\left\langle {{\vec {u}_r}|\vec {y}} \right\rangle } \left| {{\vec {u}_{\rm{r}}}} \right\rangle _2$ with
$\sum\limits_{r = 1}^N {{{\left\langle {{\vec {u}_{\rm{r}}}|\vec {y}} \right\rangle }^2}}  = 1$. According to[11], we know that $\sum\limits_{r = 1}^R {{{\left\langle {{\vec {u}_{\rm{r}}}|\vec {y}} \right\rangle }^2}}  $ should be closed to 1. So 
$p\left( 2 \right) = {\left( {\sum\limits_{r = 1}^R {\left\langle {{{\vec {u}_r}}}
 \mathrel{\left | {\vphantom {{{u_r}} \vec {y}}}
 \right. \kern-\nulldelimiterspace}
 \vec {y} \right\rangle } \frac{{{C_2}\lambda _r^2}}{{\lambda _r^2 + \alpha }}} \right)^2} \le O\left( 1 \right)$, which means we need less than $O\left( 1 \right)$ to obtain desired results on average. For  $\emph{L}$ $\alpha$, we have to perform algorithm 2 $\emph{L}$ to get all results at once. Thus, the upper bound of runtime roughly is 
$O\left( {Lpoly\log_2 M\varepsilon ^{ - 3} } \right)$. Compared with classical counterpart, we have an exponential speedup of  factors $\emph{M}$ and $\emph{N}$, whereas have slightly worse the dependence on error of a factor $\varepsilon $. Even so, our algorithm 2 have a speedup totally. In addition, there is slightly better than Yu’s improved quantum algorithm when we take same behaviour that don't take state preparing into account. Details are shown in the TABLE 2 above.
\begin{table}
\caption{Comparisons of time complexity about algorithm 2}
\begin{tabular}{cc}
\hline
\hline
Algorithm     &     Time Complexity\\
\hline
Classical counterpart & \ $
O\left( {LMN + \frac{{LN^2 \left( {\sum\nolimits_{l = 1}^K {R_l \log \left( {\frac{R}{\varepsilon }} \right)} } \right)}}{{\varepsilon ^2 }}} \right)
$\\
Yu’s algorithm & \ $
O\left( {\frac{{L\left\|{X} \right\|_{\max }^2 poly\log_2 \left( {M + N} \right)\kappa '^4 \kappa }}{{\varepsilon ^4 }}} \right)
$\\ 
Our algorithm & \ $O\left( {{{L\log_2 M} \mathord{\left/
			{\vphantom {{L\log_2 M} {\varepsilon ^3 }}} \right.
			\kern-\nulldelimiterspace} {\varepsilon ^3 }}} \right)$\\
\hline
\hline
\end{tabular}
\end{table}

As you can see from the above TABLE 2, compared with the classical counterpart, we have an exponential speedup of a factor $M$, whereas the dependence on accuracy of a factor $\varepsilon $. Liu’s algorithm doesn’t determine a good $\alpha$, so our algorithm 2 is unnecessary to compare. In addition, we are better than Yu’s improved quantum algorithm about dependence on error $\varepsilon $ and condition number $\kappa $.

\subsection{ The whole quantum algorithm for RR}
In the paper, we address RR model in quantum setting, which is one of the significant problems in data prediction. In particular, the algorithm starts with algorithm 2 to find an appropriate regularization hyperparameter $\alpha$, then we plug $\alpha$ into algorithm 1 to present a procedure that can further used to efficiently predict an output ${y'}$ for the new input $\vec x'$.

The whole time complexity is $O\left( {\left( {\kappa ^2  + L} \right)\log_2 M\varepsilon ^{ - 3} } \right)$, which has an exponential speedup over classical counterpart. 

\section{Conclusion}
\label{sec4}
 In general, we have presented a quantum algorithm that can efficiently perform quantum ridge regression over a large dataset in age of big data. Firstly, we described an quantum algorithm 1 that can predict outputs for new different input ${\vec x'}$. Secondly, we present a quantum algorithm to find a good regularization hyperparameter $\alpha$. The whole algorithm begins with algorithm 2. 
 It is shown that algorithm can address non-sparse matrices and have an exponential speedup over classical algorithm when the matrix is low-rank. Of course, there is a limitation to some certain that our algorithm only effectively tackle low-rank matrices and have slight speedup when rank of matrix is high [23].

At present, since the application of machine learning is more and more extensive and quantum algorithm is also very promising, the proposal of quantum machine learning is a very significant progress but challenging. We hope that the idea of this algorithm can be applied to other fields of information technology[24-32] to explore better possibilities in the future. What’s more, we believe that even though the field of quantum machine learning is still in an early stage of development, it will eventually an become increasingly mature field. 

\section*{Acknowledgment}
This work was supported by National Natural Science Foundation of China (Grants No. 61772134 and No. 61976053), Fujian Province Natural Science Foundation (Grant No. 2018J01776), and Program for New Century Excellent Talents in Fujian Province University.
\end{document}